\documentclass[sigconf, nonacm]{acmart}

\newcommand\vldbdoi{XX.XX/XXX.XX}

\newcommand\vldbavailabilityurl{URL_TO_YOUR_ARTIFACTS}
\newcommand\vldbpagestyle{plain} 

\begin{document}
\title{From Perils to Possibilities: \\ Understanding how Human (and AI) Biases affect Online Fora}

\author{Virginia Morini}
\email{virginia.morini@phd.unipi.it}
\orcid{}
%\authornotemark[1]
\affiliation{%
  \institution{University of Pisa}
  \streetaddress{Largo Bruno Pontecorvo, 3}
  \city{Pisa}
  \country{Italy}
}

\author{Valentina Pansanella}
\email{valentina.pansanella@unipi.it}
\orcid{}
%\authornotemark[1]
\affiliation{%
  \institution{University of Pisa}
  \streetaddress{Largo Bruno Pontecorvo, 3}
  \city{Pisa}
  \country{Italy}
}

\author{Katherine Abramski}
\email{katherine.abramski@phd.unipi.it}
\orcid{}
%\authornotemark[1]
\affiliation{%
  \institution{University of Pisa}
  \streetaddress{Largo Bruno Pontecorvo, 3}
  \city{Pisa}
  \country{Italy}
}

\author{Erica Cau}
\email{erica.cau@phd.unipi.it}
\orcid{}
%\authornotemark[1]
\affiliation{%
  \institution{University of Pisa}
  \streetaddress{Largo Bruno Pontecorvo, 3}
  \city{Pisa}
  \country{Italy}
}

\author{Andrea Failla}
\email{andrea.failla@phd.unipi.it}
\orcid{}
%\authornotemark[1]
\affiliation{%
  \institution{University of Pisa}
  \streetaddress{Largo Bruno Pontecorvo, 3}
  \city{Pisa}
  \country{Italy}
}

\author{Salvatore Citraro}
\email{{name}.{surname}@isti.cnr.it}
\author{Giulio Rossetti}
\affiliation{%
  \institution{CNR-ISTI}
  \streetaddress{Via G. Moruzzi, 1}
  \city{Pisa}
  \country{Italy}
}

%%
%% The abstract is a short summary of the work to be presented in the
%% article.
\begin{abstract}
Social media platforms are online fora where users engage in discussions, share content, and build connections.
This review explores the dynamics of social interactions, user-generated contents, and biases within the context of social media analysis (analyzing works that use the tools offered by complex network analysis and natural language processing) through the lens of three key points of view: online debates, online support, and human-AI interactions.
On the one hand, we delineate the phenomenon of online debates, where polarization, misinformation, and echo chamber formation often proliferate, driven by algorithmic biases and extreme mechanisms of homophily.
On the other hand, we explore the emergence of online support groups through users' self-disclosure and social support mechanisms.
Online debates and support mechanisms present a duality of both perils and possibilities within social media; perils of segregated communities and polarized debates, and possibilities of empathy narratives and self-help groups.
This dichotomy also extends to a third perspective: users' reliance on AI-generated content, such as the ones produced by Large Language Models, which can manifest both human biases hidden in training sets and non-human biases that emerge from their artificial neural architectures.
Analyzing interdisciplinary approaches, we aim to deepen the understanding of the complex interplay between social interactions, user-generated content, and biases within the realm of social media ecosystems.
\end{abstract}

\maketitle

%%% do not modify the following VLDB block %%
%%% VLDB block start %%%
\pagestyle{\vldbpagestyle}
%\begingroup\small\noindent\raggedright\textbf{PVLDB Reference Format:}\\
%\vldbauthors. \vldbtitle. PVLDB, \vldbvolume(\vldbissue): \vldbpages, \vldbyear.\\
%\href{https://doi.org/\vldbdoi}{doi:\vldbdoi}
%\endgroup
%\begingroup
%\renewcommand\thefootnote{}\footnote{\noindent
%This work is licensed under the Creative Commons BY-NC-ND 4.0 International License. Visit \url{https://creativecommons.org/licenses/by-nc-nd/4.0/} to view a copy of this license. For any use beyond those covered by this license, obtain permission by emailing \href{mailto:info@vldb.org}{info@vldb.org}. Copyright is held by the owner/author(s). Publication rights licensed to the VLDB Endowment. \\
%\raggedright Proceedings of the VLDB Endowment, Vol. \vldbvolume, No. \vldbissue\ %
%ISSN 2150-8097. \\
%\href{https://doi.org/\vldbdoi}{doi:\vldbdoi} \\
%}\addtocounter{footnote}{-1}\endgroup
%%% VLDB block end %%%

%%% do not modify the following VLDB block %%
%%% VLDB block start %%%
%\ifdefempty{\vldbavailabilityurl}{}{
%\vspace{.3cm}
%\begingroup\small\noindent\raggedright\textbf{PVLDB Artifact Availability:}\\
%The source code, data, and/or other artifacts have been made available at \url{\vldbavailabilityurl}.
%\endgroup
%}
%%% VLDB block end %%%

\section{Introduction}
%\section{Models, tools and biases}
Online social media platforms (henceforth, OSPs) are virtual spaces where individuals freely share their thoughts, emotions, and opinions with a wide audience of other users.
Due to their pervasiveness in our daily lives, actions carried out on OSPs (e.g., information production/consumption, debates, self-disclosure\dots) also reverberate in the physical world, deeply affecting the way we form our opinions, build interpersonal ties, and share personal narratives.

While offering unprecedented opportunities, such online fora are not neutral.
Human biases - e.g., our tendency to seek confirmation of our ideas, to surround ourselves with like-minded peers, to selectively distance opposing narratives \dots - are often amplified by OSPs due to the (pseudo)anonymity online users can shield themselves with, sometimes leading to phenomena that can be detrimental for the individual and society as a whole.
Moreover, since such environments are proper socio-technical systems, those biases are also extended and reinforced by those introduced by algorithmic data curation - e.g., recommendations (involving suggested interactions with both users and contents) and representation biases (involving how AI models generate synthetic, human-like contents).
In this complex scenario, in which it is unclear whether a balance can exist between possibilities and perils, multidisciplinary research is called upon to unveil those relevant patterns emerging from digital traces and to assess their effects on the users of OSPs.

Multidisciplinary approaches are needed to study such rapidly evolving realities.
Recent literature has seen the development of theoretical and analytical studies, encompassing quantitative and qualitative research, targeting different aspects of such phenomena.
Online users' interactions are often studied through social network analysis modeling lenses to unveil topological patterns and explain social dynamics observed in such settings.
Similarly, the contents they produce and consume have been extensively analyzed to measure their emotional values, identify discussion contents, and compare the virality of legitimate and misleading news.
All such research endeavors, primarily in STEM, are often complemented and guided by sociological and psychological theories to obtain local validations of global behaviors.

In this survey paper, we discuss recent work addressing such a complex research area, focusing on three different but interrelated topics - namely, online debate pollution, online support seeking, and conversational AI. We highlight the potential effects of human and AI biases, discussing how quantitative and qualitative research efforts are crucial for better understanding OSPs' pressing issues.
The paper is organized as follows: in Section \ref{sec:debates} we revise the realm of online debates, discussing how homophily and related phenomena might impact opinion dynamics, sometimes leading to the formation of epistemic enclaves; Section \ref{sec:support} discusses the impact of peer-pressure and social support on the emotional journeys of online users that leverage the pseudo-anonymity of OSPs to self-disclose personal issues; Section \ref{sec:LLM} shifts the focus to conversational AI systems, discussing their biases, cognitive limitations, and the potential effects they have on online human-AI interaction.
Finally, Section \ref{sec:conclusion} concludes our review, providing a discussion of potentially valuable future research lines.

\section{Online Debates: Pollution and Biases}
\label{sec:debates}
Online debates often anticipate/extend those polarized social interactions that can be observed in the physical world. 
Many controversial cases have already highlighted the effects that coordinated actions - born and consolidated online - have on our society: from obstruction to vaccination deployment \cite{cossard2020falling} to inference on political elections \cite{bovet2019influence}, ethnic stigmatization \cite{budhwani2020creating}, and post-electoral riots (e.g., in the US and Brazil).
To characterize such online-to-offline events, research efforts focused on analyzing those phenomena that, allegedly, play a role in polluting online debates.
Multidisciplinary research activities have been carried out to discriminate trustable from untrustable contents \cite{firoj2022survey}, to identify bots and their coordinated activities \cite{caldarelli2020role}, to assess the presence of d/misinformation campaigns \cite{hernon1995disinformation}, propaganda \cite{nakov2020fact}, and identify Echo Chambers (henceforth, ECs) \cite{cinelli2021echo, garimella2018quantifying, ge2020understanding, hilbert2018communicating}. %de2021no
Moreover, to understand the potential effects of polluted debates, opinion dynamics literature \cite{sirbu2017opinion} witnessed the definition of models devised to describe how individuals exchange opinions online. 
Those models - accounting for polluting factors such as response to fake news \cite{toccaceli2021opinion}, the presence of stubborn agents/polarized media, and recommendation systems selection bias \cite{sirbu2019algorithmic,pansanella2023mass} - explicitly aim at studying how opinions and beliefs vary over time.
Indeed, when talking about ``pollution'', we are referring to an ensemble of deviant phenomena that have often been studied independently one from the other, although their effects/existence are allegedly strongly intercorrelated. 
In the following, we delve into a discussion of three intertwined research lines focusing on modeling and measuring biases (and their effects) on online (polluted) debates: Homophilic behaviors estimation, Opinion Dynamic modeling, and Echo Chamber analysis.
\\ \ \\
\noindent{\bf Homophily and Acrophily: a path toward Radicalization.}
\label{sub:homo}
Homophily is a basic organizing principle that refers to the tendency of individuals to associate and bond with others sharing similar beliefs, values, and demographics~\cite{mcpherson2001birds}.
This idea that \textit{similarity breeds connection} has been observed in a wide range of contexts, both in the offline~\cite{newman2003mixing} 
and online domains~\cite{barbera2015tweeting}.
Online – and on social media in particular – this property was linked with the emergence of polarized environments%~\cite{barbera2015tweeting} , cinelli2021echo} 
and claimed to be fueled by both self-selection~\cite{barbera2015tweeting} and algorithmic choice~\cite{kordzadeh2022algorithmic}. %, pariser2011filter}.
Traditionally, homophily has been measured with Newman's nominal assortativity coefficient~\cite{newman2003mixing}, a summary statistic computing how interactions between nodes of the same group deviate from the overall group mixing. 
Despite its popularity, this measure suffers from known limitations, such as sensitivity to group size imbalance and globality – an undesired property in large systems, in which varying degrees of homo/heterophilic mixing may coexist~\cite{peel2018multiscale}. %, karimi2023inadequacy}.
To overcome these issues, research has moved toward new indicators independent of group sizes~\cite{karimi2023inadequacy}, and/or focus on local homophily estimation~\cite{rossetti2021conformity}.
This latter research line has also explored local time-aware measures~\cite{citraro2022delta}, and higher-order interactions~\cite{failla2023attributed}.

Recent research stemming from social psychology has uncovered another tie-formation mechanism closely tied to homophily, namely \textit{acrophily}.
Acrophily identifies the tendency to affiliate with like-minded peers having more extreme (rather than moderate) views and refers especially to the political domain~\cite{goldenberg2023homophily}.
A seminal study of acrophily on Twitter/X finds that acrophilic users are more likely to use plural pronouns (e.g., \textit{us}, \textit{them}) and post negative content~\cite{Goldenberg2023attraction}. 
Moreover, evidence of the joint action of homophily and acrophily shows that such a combination eventually leads to higher segregation than the effect of homophily alone. %~\cite{Goldenberg2023attraction}.
It is thus self-evident that the interplay of these phenomena can lead to the progressive extremization of opinions, ultimately cascading into radicalization.
Indeed, these phenomena become key dangerous factors when identifying experts in online debates, as they likely facilitate preference for extreme, like-minded individuals.
This may result in excessive fragmentation, stubbornness, and opinion stagnation~\cite{sunstein2018republic, nguyen2020cognitive}.
Radicalization is framed by multiple theories~\cite{moghaddam2005staircase, van2014going} %, schmid2013radicalisation, neo2019internet} 
as centering around (i) an \textit{us vs. them} scenario, which is a natural side-effect of homophily and allegedly common under acrophily~\cite{Goldenberg2023attraction}, and (ii) a \textit{journey} toward extreme views, which acrophilic tendencies might ease.
However, data-driven research on homophily and acrophily and their implications for radicalization is still limited.
\\ \ \\
\noindent{\bf Opinion Dynamics: Polarization and Fragmentation.}
\label{sub:opinion}
Opinions wield considerable influence in shaping individual and collective behavior across various domains. 
Social influence theory highlights the role of interactions in driving opinion evolution, leading to a convergence of viewpoints over time. Social networks, serving as conduits for interactions, perpetuate biases and homophily tendencies, shaping opinion formation.
Traditional and digital media influence opinions, yet biases inherent in media coverage can distort public discourse. 
In the digital age, online platforms and algorithms mediate information consumption, exacerbating confirmation bias and filter bubbles and limiting exposure to diverse perspectives.
Internal cognitive biases significantly shape opinions, with psychological factors often leading individuals to reject opposing viewpoints and reinforce existing beliefs.
\\ 
\noindent{\it What is an opinion dynamics model?}
When we talk about opinion dynamics models - also called mathematical models of social learning - we refer to a wide variety of studies and approaches that aim at understanding how opinions form and evolve through social interaction processes \cite{castellano2009statistical}.
Models of opinion dynamics generally consider a population of individuals and numerically simulate the interactions between them - whenever possible, they compute the final state analytically. 
Rules normally govern such processes %- often even very simple equations - 
developed according to empirically observed sociological behaviors, chosen by the scientists to reproduce patterns observed in the real world and provide a causal explanation of them.
%Among them, the most frequent modeled biases are the cognitive, algorithmic and external ones.
One of the most famous and earliest attempts at capturing possible drivers of polarization is the introduction of \textit{bounded confidence} in opinion dynamics models. 
The so-called \textit{Bounded Confidence} models constitute a broad family of models where agents are influenced only by peers having an opinion sufficiently close to theirs, namely below a certain confidence threshold -a behavior grounded in \textit{homophily}-related sociological theories. 
Such models have been used to replicate individuals' lack of understanding, conflicts of interest, or social pressure - thus capturing cognitive biases \cite{deffuant2001mixing}.
%This threshold has also been referred to as open-mindedness \cite{lorenz2010heterogeneous}, while some may argue that it is more similar to influenciability. 
%\paragraph{Algorithmic Biases}
The role of algorithmic bias in the process of opinion evolution has been extensively studied, with models going from single-parameter toy recommender systems  \cite{sirbu2019algorithmic} to synthetic replicas of state-of-the-art models.%\cite{cinus2022effect}.
Even a simple model such as \cite{sirbu2019algorithmic} is able to capture the {\emph fragmenting} power of algorithmic biases, showing how systematically skewing interactions towards like-minded individuals fosters polarization and fragmentation. 
This is evident in three dimensions: a higher number of opinion clusters, a higher distance between opinion clusters, and a higher time needed for agents to cluster around fixed opinions. 
These results hold across different modeling frameworks and/or network topologies used to proxy the underlying social network \cite{pansanella2022mean} %, peralta2021effect, peralta2022opinion}, 
with the additional element of sparser connection, which worsens the final fragmentation of opinions \cite{pansanella2022mean}. 
%The situation at the level of polarization and segregation of the population into clusters of opposing views is further exacerbated when we consider the ability of users to break and reform connections in ways that minimize conflicting interactions \cite{Kan2021AnAB, Kozma2008ConsensusFO, pansanella2022modeling}. 
%The presence of algorithmic bias then makes the network structurally less segregated in this case since it  minimizes conflicting interactions by preventing users from ``seeing" their discordant connections) \cite{pansanella2022modeling}. 
\cite{maes2015will} argued that personalization algorithms could either foster or weaken polarization, depending on assumptions about how opinions are reinforced or rejected through interactions. 
However, empirical studies suggest that while recommender systems may increase polarization according to audits and models, real data indicates that user choices play a significant role, often avoiding niche or extremist content. 
This apparent paradox underscores the need for nuanced interpretations of algorithmic amplification. 
Additionally, ecosystem characteristics can independently contribute to the formation of echo chambers. 
%Overall, the complexity of modeling personalization algorithms arises from limited understanding of their real-world mechanisms, emphasizing the ongoing challenge in assessing their impact.
%\paragraph{External Effects}
%- \cite{sirbu2017opinion}
%- \cite{pansanella2023mass}
\\ \ \\
\noindent{\bf Echo Chambers: Social filters to Information pluralism.}
\label{sub:EC}
One of the subjects that attracted increased attention is the tendency of polarized users to create Echo Chambers \cite{pariser2011filter}: social structures whose members agree on a predefined set of values and tend to discredit alternative ones. 
ECs are epistemic structures that, once established a tacit knowledge \cite{collins2007bicycling, lynch2013margins}, support their members by enforcing an in/out-group dialectic (e.g., communities of practice or interest \cite{ardichvili2003motivation}) which is functional to feed confirmation bias.
Nguyen \cite{nguyen2020echo, nguyen2020cognitive} discusses some peculiarities that differentiate echo chambers from Epistemic Bubbles, underlying how the latter represent loose structures whose members are not yet polarized, rather unable to access a balanced information diet. 
Counterintuitively, information diffusion processes are not blocked by echo chambers \cite{santos2021echo}: their members are aware of the alternative opinions and information living outside their in-group but decide not to accept them and distance further whenever possible by enforcing the so-called backfire effect (a first, although not sufficient per se \cite{nyhan2021backfire}, mechanism to defend from opposite views). 
Conversely, Epistemic Bubbles are digital enclaves where alternative information/stances are kept out through gatekeeping \cite{baumgaertner2022preference}. 
Nguyen postulates that escaping from echo chambers is a rather complex task requiring a cognitive reset while leaving an Epistemic Bubble is only a matter of losing the user's social homophilic strength. 
Such a theoretical framework is a clear example of how different pollution phenomena interplay: social selection, m/disinformation access reinforced by gatekeeping, and opinion polarization are different sides of the same coordinated behaviors that affect individuals' online whereabouts. 
Moreover, due to such intertwined effects, newcomer users to a polluted debate are at risk of being dynamically trapped \cite{begby2022belief} into cognitive islands \cite{nguyen2020cognitive}, local epistemic realities that %- due to the impossibility of newcomers to identify experts - 
might lead to individual runaway echo chambers: self-feeding processes that incrementally tend toward radicalization of opinions \cite{banisch2022modelling}.
Several works attempted to identify and measure echo chambers in social media, leveraging both SNA and NLP techniques and focusing on different topological scales \cite{cinelli2021echo,morini2021toward,buongiovanni2022will}.
Recently, rising attention has been placed on capturing and tracking the evolution of such epistemic realities to understand their stability and assess the risk of online users getting trapped in them due to confirmation, selection, algorithmic, and cognitive biases.

\section{Online Support: Narratives and Personal Disclosure}
\label{sec:support}
Up to this point, our discussion has centered on how social media platforms can amplify users' cognitive biases when engaging in debates over controversial and divisive topics, potentially leading to concerning outcomes such as opinion polarization or the formation of echo chambers. 
However, when observed in other contexts, echo chamber-like structures may not be inherently detrimental and could yield positive outcomes for individuals and society. 
Online homophilic groups often form amongst people facing challenges, such as a particular illness or a stigmatized issue, facilitated by the on-demand ease of communication and the possibility of being anonymous  \cite{torous2016role}.
Accordingly, in the last decade, we have assisted in the rise and growth of \textit{online self-help groups} (OSHGs), i.e., peer support networks of mutual giving and receiving where individuals feel free to share their personal experiences with ``sympathetic others", seek information about specific issues, and give and receive support from others facing the same social stigma \cite{bucci2019digital}. %, goffman2009stigma, berry2017whywetweetmh}. 
Data-driven studies exploring supportive online discourse mainly focus on different kinds of OSHGs (e.g., mental health issues, addictions, health concerns) observed on Reddit, X/Twitter, and Facebook.
The impact of online supportive discourse has been usually framed in the light of \textit{social support} and \textit{self-disclosure} phenomena. 
\\ \ \\
\noindent{\bf Self-disclosure.} \label{sec:selfdisclosure}
The phenomenon of self-disclosure - also called ``willful disclosure'' - is the act of revealing personal information to others in such a way as to express feelings, develop trust, and build intimacy with others \cite{cozby1973self}. 
Traditionally, disclosing personal information, fears, and experiences has been an established phenomenon in psychotherapy through the therapist and client relationship. 
In contrast to such dyadic disclosures, online platforms have become arenas for ``broadcasting self-disclosures," i.e., sharing personal information in public contexts, often to invisible audiences, and supported by the tips of anonymity or semi-anonymity of these platforms. 
A widely used approach to recognize and study attempts of online self-disclosure is to analyze textual contents through \textit{Open coding} i.e., by recruiting a set of researchers who, given a sample of posts on a specific issue, carefully analyze them and extract insights according to known Behavioral frameworks. 
For instance, in \cite{andalibi2018social} %,andalibi2016understanding
 authors focus on sexual abuse-related disclosures on Reddit by analyzing the different writing behaviors between identified and ``throwaway" (i.e., anonymous) accounts. 
In detail, they manually identify different levels of disclosure by referring to the Disclosure Processes Model \cite{chaudoir2010disclosure} that helps in understanding when and why interpersonal disclosure may be beneficial in stigmatized contexts. 
Such a study reveals that anonymity plays a critical role in support-seeking and disclosure behaviors: posts made using throwaway accounts show distinct content from those made by identified accounts, indicating the importance of anonymity for survivors seeking support.
In addition, \cite{andalibi2017sensitive} leverages Bernard Rimé's framework of \textit{social sharing of emotion as meaning creation} \cite{rime2009emotion} to explore self-disclosure across Instagram posts that contain the hashtag \#depression, suggesting that self-disclosures are not merely about sharing personal experiences but mainly about engaging with an audience. 
Users employ strategies such as direct questions, calls for support, and sharing personal narratives to invite interaction and show awareness of their audience. 
Unlike such qualitative methods, quantitative approaches search for online self-disclosure by automatically extracting different textual features that can reveal the emotions, sentiments, and topics expressed from shared posts. 
In the literature, there exists a very broad set of tools/algorithms (e.g., LIWC\footnote{https://www.liwc.app/}, TFMNs \cite{stella2020text}, BERTopic\footnote{https://github.com/MaartenGr/BERTopic}, Empath \cite{fast2016empath}), and lexicons (e.g., NRCLex \cite{mohammad2013crowdsourcing},    VADER \cite{hutto2014vader})  that can help researchers detect users' emotional states based on what they write and share on OSPs. 
For instance, \cite{ernala2017linguistic} quantitatively investigates behavioral changes preceding and following disclosures of schizophrenia on Twitter, underlying that, after disclosure, users-generated content shows increased positive affect, enhanced readability, improved topical coherence, future orientation, and reduced discussion on schizophrenia symptoms and stigma. 
In \cite{chen2021exploring}, authors examine the effects of received empathy on users' engagement with OSHGs, specifically looking at users' probability of returning and their subsequent expression of empathy within these communities. 
They uncover that receiving social support at the first post significantly increases users' likelihood of returning to OSHGs and their subsequent expression of empathy.
Moreover, in this scenario, capturing a topical representation of the issue discussed could be useful without relying on an open coding process. 
To such an extent, topic modeling algorithms were successfully used to extract topic information from different Reddit OSHGs \cite{morini2023can}. %chen2021exploring
Another tool useful for extracting knowledge and emotions from text is Textual Forma Mentis Networks (TFMNs), a type of cognitive network that transforms text into complex network representations. 
In these representations, nodes represent concepts, and edges represent syntactic and semantic relationships.
TFMNs can be enriched with data from external lexicons to facilitate in-depth quantitative and qualitative analyses of the thoughts and emotions expressed by users as they relate to certain topics. 
In \cite{abramski2023voices} TFMNs are applied to study the perspectives of rape survivors as expressed in online personal narratives on Reddit about passive and active voice usage - showing that rape survivors who describe their experiences using the passive voice place a greater focus on concepts related to psychological distress (PTSD, anxiety, trauma), while those who describe their experiences using the active voice focus more on concepts related to family members (parent, father, brother).

Similarly, \cite{joseph2021cognitive} investigates the thoughts and feelings expressed in suicide notes, underlying that they are characterized by a balanced compartmentalization of positively and negatively valenced concepts, shedding light on the complex nature of the thoughts and emotions possessed by those who choose to take their own lives.
\\ \ \\
\noindent{\bf Social Support.}
The anthropologist Berton Kaplan defines \emph{social support} as the ``degree to which an individual's needs for affection, approval, belonging, and security are met by significant others" \cite{kaplan1977social}. 
It is recognized that individuals facing particular challenges benefit from interactions with and support from ``sympathetic others" who share the same social stigma and have had similar experiences \cite{goffman2009stigma}.  
According to the traditional categorization schema on social support - Social Support Behavioral Code (SSBC) \cite{suhr2004social} - there exist different types of support that an individual can offer: informational support that consists of providing information about the stress event or how to deal with it; instrumental or tangible support, i.e., offering goods or services needed in a challenging situation; emotional support that means offering love or caring and network support, i.e., giving companionship to an individual facing difficulties. 
As for self-disclosure, a common way to assess if and to what extent a post receives social support is to rely on the open coding procedure. 
In this scenario, researchers tend to categorize the comments that a post receives according to the SSBC framework. 
This is the case for \cite{andalibi2018social}, focusing on social feedback in response to sexual abuse disclosures on Reddit, and for \cite{andalibi2017sensitive}, where replies on Instagram posts talking about depression are categorized concerning the informational and emotional support received. 
Interestingly, they observed that certain types of content, such as references to illness and concerns about self-appearance, are more likely to attract positive social support.
Another widely and straightforward proxy used to quantify how much support a post/user receives is to count its number of comments \cite{de2017language}.
%Several works, e.g.,  \cite{de2017language} %,cunha2016effect,de2014mental} 
% consider commentary in OSHGs as a mechanism through which support is extended to help seekers.  
To understand if online social support has a real effect on individuals and, thus, therapeutic outcomes, research focused on capturing if users are more prone to write again on OSHGs after receiving help - designing a control/treatment study approach and estimating the effect of receiving or not receiving social support.
%Accordingly, observed population is splitted into a control and treatment group. 
%The first comprises all users that receive advice/help/support to their first post; the second represents those without replies. 
%Then, the subsequent step consists of estimating the effect of receiving or not receiving social support. 
The majority of studies \cite{chen2021exploring,de2017language} have adopted the statistical technique of \textit{propensity score matching} (PSM). 
Indeed, PSM attempts to estimate the effect of a treatment, policy, or other intervention by accounting for the covariates that predict receiving the treatment. 
Chen et al. \cite{chen2021exploring} discovered that social support and empathy are ``contagious" within OSHGs: users who received social support or empathetic responses in their first posts were more likely to return to the communities, post again, and offer support to others. 
Particularly, \cite{de2017language} shows that users who benefit from online social support tend to show greater social and futuristic orientation, interpersonal awareness, and lower cognitive impairment. 
%Authors in \cite{balsamo2023pursuit} leverage  \textit{Interrupted Time Series} analysis to investigate the potential of the Reddit community in providing peer support to individuals during the initial stages of opioid use recovery. 
%Again, they found that supportive behaviours within the community encourage users to abandon opioid-related communities in favor of recovery-oriented ones. \newline 
Only a few works analyzed the impact of supportive dynamics in OSHGs by combining features extracted from user-generated content and the underlying social network of user interactions.
In \cite{morini2023can} the social effects of discussion and interactions on depression-related Reddit communities are analyzed. 
First, four distinct archetypal profiles that resonate with the Patient Health Engagement model \cite{graffigna2016patient} were identified through user-generated content. 
Then, by analyzing users' interaction patterns with different profiles, the authors unveil that users' profile transitions follow non-linear journeys among these archetypes,  navigating through both positive and negative phases in a spiraling rather than a linear progression.

\section{From Human Biases to LLM ones}
\label{sec:LLM}
The rise of Large Language Models (LLMs) has prompted the need to assess how AI performance aligns with human cognitive functions, ranging from decision-making to language comprehension and generation \cite{mitchell2023debate}.
This exploration is crucial for understanding the expressive power of LLMs, recognizing their limitations, and establishing boundaries around human reliance on LLM-generated content.
In the context of generative LLMs, the risk of blindly trusting AI output is considerably high, particularly when these models fail to recognize the human-like bias they inherit \cite{acerbi2023large}.
Despite differing perspectives on the nature of intelligence demonstrated by LLMs, the harmful biases exhibited by LLMs has emerged as a ubiquitous concern. 
Biases, stemming from the under-representation of data to the over-representation of societal stereotypes in training sets, could potentially worsen discrimination and societal inequalities, posing significant challenges in user interactions with LLMs. 
At the core of this phenomenon is the causal embedding hypothesis \cite{caliskan2020social}, which states that biases can be learned from exposure to language statistics that reflect social stereotypes. %caliskan2017semantics, 
%This is true of both machines and humans. 
%For example, we may acquire implicit attitudes about scientists and gender through exposure to language in which the word \textit{scientist} occurs more frequently with male-gendered words like \textit{he} and \textit{his}. 
%Such exposure can lead to the acquisition of implicit attitudes that reflect the belief that men make better scientists than women, despite never having been exposed to such an explicit proposition. 
The understanding that many biases originate in language has led to a surge in research investigating biases in language models. 
For example, studies have found that even the most advanced models posses significant racial and gender biases \cite{manzini2019black, lucy2021gender},  %like father is to doctor as "mother" is to "nurse" 
%\cite{bolukbasi2016man}, and black is to criminal as Caucasian is to police \cite{manzini2019black}. 
%and showed that GPT-3 (Generative Pretrained Transformers)-generated-stories exhibit gender stereotypes such that feminine characters are associated with family topics and masculine characters are described with higher power verbs \cite{lucy2021gender}.
%Even vanilla implementations of GPT-like models inherit subtle biases such as implicit negative connotations toward math and STEM subjects \cite{abramski2023cognitive}. 
%These studies demonstrate how even the most advanced language models can display biases similar to our own.
Such studies demonstrate the need to understand the nature of biases in LLMs as they relate to human biases, in order to mitigate the potentially harmful effects of widespread LLM interaction.

%However, despite being trained on human data, researchers have begun questioning whether LLMs can also manifest non-human biases \cite{stella2023using}.
%The concern about non-human bias emerges because LLMs do not learn as humans.
%The knowledge LLMs possess is not innate \cite{berwick2016only}, embodied (e.g., linked to sensorimotor experiences) \cite{varela2017embodied}, or guided by metacognition, the ability to understand one's mental process \cite{Premack1978}.
%Myopic overconfidence and hallucinations have been described as non-human biases that characterize LLMs as systems failing to make inferences about missing data and their data sources \cite{stella2023using}.
%This could be interpreted as a lack of innate, embodied, metacognitive knowledge, prompting LLMs to compensate for gaps in their training sets, producing non-human bias.
In the following, we present a brief review of emerging lines of research with the aim to better understand the nature of LLM biases, and in particular, whether LLMs can easily replicate human-specific cognitive functions.
These studies adopt the tools of network science e.g., analyzing the underlying network structure of word representations, or employing tools from opinion dynamics.
\\ \ \\
\noindent{\bf Representation Bias.} Semantic representation refers to how the meaning of words is represented in a structured way. 
The biased outputs produced by LLMs originate in the biased semantic representations they possess. 
These biased representations can be investigated using several approaches, depending on the architecture of the language model. 
Earlier language models that relied on static embeddings (a single vector representation for each word) could be investigated by directly accessing their embedding spaces \cite{caliskan2020social}.%, bolukbasi2016man, manzini2019black, caliskan2017semantics, }.
Instead, for newer models that rely on contextual embeddings (a unique vector for each instance of a word), investigating the embedding space poses many problems \cite{apidianaki2022word}. 
These challenges have led to a shift from the traditional bottom-up approach to investigating language models in favor of a top-down approach that probes LLMs in various tasks, using their output to understand the nature of their biased representations. 
This top-down approach can be compared to cognitive psychology approaches that aim to understand how humans think using data from behavioral experiments. 
%In fact, theories and methods for investigating how humans represent concepts can be applied to LLMs in order to understand the nature of their biases, including understanding how they are similar and different from human biases.
Investigations of the mental lexicon, which aim to understand how humans represent concepts in terms of relationships between them, can be particularly useful for understanding biases in LLMs. 
Models of the mental lexicon, also called semantic memory, can be constructed in many ways, for instance, by leveraging network science modeling. Cognitive network models of semantic memory represent concepts as nodes and relationships between them as edges. Decisions about how to construct the networks reflect hypotheses about how knowledge is structured. For example, a knowledge-based approach attempts to reconstruct semantic memory \textit{a priori} based on our understanding of the knowledge we possess, forming connections based on hierarchical relationships or relations derived from dictionary definitions. Instead, behavior-based approaches attempt to reconstruct semantic memory \textit{a posteriori} based on data observed in behavioral experiments, such as the free association task, which prompts participants to produce three words that come to mind when given a cue word. Network models built from such data have been extensively applied to study a wide variety of cognitive phenomena such as language learning \cite{citraro2023feature} and creativity \cite{benedek2017semantic}
Recent work in this area has demonstrated that multiplex, higher-order and feature-rich network models \cite{citraro2023hypergraph, citraro2023feature} can provide further insights about complex cognitive processes. Just as models of the mental lexicon can be used to study human cognition, they may be applied for studying LLMs by constructing networks using LLM-generated data rather than human-generated data. Representation biases may be investigated in these models by exploring the network properties related to specific concepts within the networks. Such an approach was applied to investigate LLM biases related to math and STEM subjects \cite{abramski2023cognitive}. This direction of research offers exciting opportunities for investigating the similarities and differences between human and LLM biases using an approach that is flexible and easily interpretable.
\\ \ \\
\noindent{\bf Theory of Mind}
%- Opinion dynamics LLM agents
(ToM) \cite{Premack1978} is a concept in cognitive psychology that refers to all those processes that allow the human mind to attribute mental states to itself and others \cite{Cuzzolin2020}. 
This internal reasoning about the mental states of external people forms a kind of \textit{meta-representation} \cite{Perner1993} of others' internal states, which includes understanding and even predicting their future behaviors, beliefs, and desires \cite{Papera2019}. 
In addition, this ability allows someone with a well-developed ToM to understand that someone else may have a belief about a third person's thoughts or intentions, resulting in a \textit{second-order} representation.
%Since the publication of the original paper, much attention has been drawn to this area (e.g., \cite{Wellman2001, Apperly2012}) due to the importance of ToM for social interaction and language.
Lack of ToM is believed to be implicated in poor comprehension of social situations and lack of empathy; furthermore, lack of ToM might even be the cause behind the difficulties experienced by people on the autism spectrum in understanding the mental states of others \cite{BaronCohen2000}. %TagerFlusberg1992
Recent research is trying to determine whether LLMs need to learn a ToM or, more intriguingly, whether they have already learned it from the huge amount of training data. 
Early works \cite{Cuzzolin2020} discussed and highlighted the need for AI to tackle \textit{hot cognition} \cite{Abelson1965}, especially the ToM, and propose developing a \textit{Machine Theory of Mind}, using reinforcement learning and neural networks to handle the connections between mental states .%- arguing for a closer collaboration between computer scientists, psychiatrists, and psychologists.
Currently, ToM in LLMs is assessed by employing traditional batteries of tests already used for children. 
For instance, \cite{Kosinski2023} presented evidence that GPT models can solve traditional false-belief tasks, leading the authors to hypothesize that ToM may develop as an emerging behavior in LLMs during the training phase, similar to the case of the ability to count \cite{Nasr2019} (which emerged spontaneously in image processing models).
%The author also suggests that other factors may be involved in this result, such as recalling similar stories encountered in the training data.
A similar result was obtained by \cite{Brunet2023}, where ChatGPT was exposed to several ToM tasks: such an outcome can be related to the model using a particular character's belief or mental concepts, provided it is under certain prompting conditions.
%Additionally, Guo \textit{et al.} \cite{Guo2023} improved GPT-4 abilities in solving an incomplete information game by designing specific prompts to inject a ToM.
%\\
In contrast, \cite{Ullman2023} challenges such results by showing that even small perturbations in the task may reverse the results and affect the ToM abilities of LLMs. 
Moreover, \cite{Sap2023} refuted LLM ToM theory, raised skepticism about literature results, and showed how GPT-3.5 and GPT-4 fail to probe the mental states of others -- even if the LLM shows more acceptable skills in answering factual questions. 
Discussing ToM in LLM is crucial due to their increasing relevance as a component of socio-technical systems.
%--- Perché è importante ToM? socio-technical systems!!
\\ \ \\
\noindent{\bf Biases: from Humans to LLMs and back.}
The adoption of synthetic agents (e.g., LLMs) as proxies for human behavior is a growing trend. 
If we assume a ToM for synthetic agents, a new frontier in social simulations emerges, designing a playground where agents can act based on past experiences and respond realistically to their environment.
Preliminary studies reveal that LLMs populations can manifest emerging behavior similar to human societies \cite{leng2023llm}, e.g., the formation of scale-free networks \cite{de2023emergence}, information diffusion \cite{gao2023s}, trust \cite{xie2024can}.% park2023generative - subrational behaviors \cite{coletta2024llm}, quarantining in epidemics settings \cite{williams2023epidemic} and others.
Considering opinion evolution simulation, initial results showed that hybrid frameworks leveraging LLMs agents outperformed pure agent-based models (ABMs) in replicating echo chamber behavior in Twitter-like environments \cite{mou2024unveiling}.
Introducing confirmation bias through prompt engineering can lead to opinion fragmentation in LLM-agents populations \cite{chuang2023simulating}, in line with existing agent-based modeling and opinion dynamics research; polarization also emerges when LLM-agents are in an echo chamber due to the model's ability to understand prompts and update its own opinion by considering both its own and the surrounding opinions as well as the persona assigned to the agent. 
LLM-powered agents can also resemble dynamics of persuasion and opinion change typically observed in the human discourse \cite{breum2023persuasive}, e.g., generating persuasive arguments incorporating social pragmatics supporting psycho-linguistic theories of opinion change \cite{monti2022language}. 
However, \cite{flamino2024limits} found that LLM agents are less convincing in debate than humans. 
Unlike human agents, if not properly controlled, LLM-powered agents have an inherent bias towards producing accurate information, leading populations of such agents towards consensus in line with scientific reality \cite{chuang2023simulating}, or towards an inherent bias of the model \cite{taubenfeld2024systematic} even in an echo chamber environments. 
Moreover, LLMs are also biased toward being more polite, articulated, and respectful than users on real-world social media platforms \cite{tornberg2023simulating} and may be biased in their ability to represent different populations \cite{santurkar2023opinions}.%, prompting the need to train and calibrate LLMs using data sourced from existing platforms, replicating the behavior of each individual in real-world settings. 
%Furthermore, the results of such simulations need to be carefully controlled; even if results resemble collective behavior, the reasons behind this might not be the same. For example, in \cite{de2023emergence}, the emergence of scale-freeness could also be due to an inherent bias of the used LLM towards certain agent names rather than preferential attachment. 
Therefore, while Large Language Models (LLMs) offer promising avenues for simulating complex social dynamics, their inherent biases and limitations necessitate careful calibration and control to accurately replicate human behavior and avoid oversimplified conclusions from simulation results.
Moreover, when embedded into socio-technical systems, human biases, and LLM ones might feedback loop on each other, thus producing unexpected and unpredictable emerging behaviors.

\section{Conclusions}
\label{sec:conclusion}
Online interactions involving solely human actors or extending to AI agents are nowadays relevant pathways through which information diffuses and opinions form and develop.
In this paper, by focusing on three disjoint although related environments, we discussed how such dynamics might open for unprecedented possibilities - e.g., supporting personal growth, favoring opinion exchange, augmenting users' knowledge - while, at the same time, raising concerns about perils tied to human/AI-related bias amplification. 
Our analysis showed how biases might affect users discussing global themes as well as personal development; moreover, we also debated on the limitation of LLMs' ability to understand their users' means and intentions and discussed how that, along with their own emerging biases, might end up affecting their use as ``smart agents" to perform social simulations.
Multidisciplinary, data-driven theories and studies targeting such themes are needed today more than ever due to the direct effect the online experience has on individual users and society.
Among the most overlooked analytical dimensions opening up for valuable research are temporal analysis and user-system co-evolution, which are relevant challenges to be tackled.
So far, data-driven studies tend to focus on crystallized realities - snapshots of complex evolving phenomena - thus strongly reducing the impacts of their observation.
Online debates, self-disclosure, and AI-mediated interactions are dynamic in nature, and neglecting such aspects can lead to misleading interpretations while, at the same time, reducing the possible insights on phenomena evolution that can support (whenever needed) mitigation efforts.
Similarly, users' behaviors (and biases) feedback loop into the AI ones - e.g., in recommender systems retraining: disentangling the effects of such a complex dynamic might open up novel socio-technical system designs tailored to support users' online journeys, providing debiased, safer environments.

\begin{acks}
This work is supported by (i) the European Union – Horizon 2020 Program under the scheme “INFRAIA-01-2018-2019 – Integrating Activities for Advanced Communities”, Grant Agreement n.871042, “SoBigData++: European Integrated Infrastructure for Social Mining and Big Data Analytics” (\url{http://www.sobigdata.eu}); (ii) SoBigData.it
which receives funding from the European Union – NextGenerationEU – National Recovery and Resilience Plan (Piano Nazionale di Ripresa e Resilienza, PNRR) – Project: “SoBigData.it – Strengthening the Italian RI for Social Mining and Big Data Analytics” – Prot. IR0000013 – Avviso n. 3264 del 28/12/2021; (iii) EU NextGenerationEU programme under the funding schemes PNRR-PE-AI FAIR (Future Artificial Intelligence Research).
\end{acks}

%\clearpage

\bibliographystyle{ACM-Reference-Format}
\bibliography{sample}

\end{document}